\documentclass[a4paper,twoside]{article}

\usepackage{fancyhdr}
\usepackage{multicol}
\usepackage{upgreek}
\usepackage{indentfirst}
\usepackage{booktabs}
\usepackage{array,tabularx}
\usepackage{keyval,graphicx}
\usepackage{textcomp}
\usepackage{amssymb,bm,mathrsfs,bbm,amscd}
\usepackage[tbtags]{amsmath}
\usepackage{lastpage}
\usepackage[numbers,sort&compress]{natbib}  

\newcommand{\newsmall}{\fontsize{9pt}{0.8\baselineskip}\selectfont}

\DeclareMathSizes{10}{10}{6}{5}

\font\scten=euex10 at 10pt
\newcommand{\vint}{\mathop{{\vcenter{\hbox{\scten\char90}}}}}

\newcommand{\dao}{{\rm d}}

\newcommand{\emark}[1]{$^{#1}$}

\newcommand{\thanksmark}{\textsuperscript{\,\rm{*}}}

\renewcommand{\,}{\hspace{0.125em plus 0.025em minus 0.025em}}

\newcommand{\ruledown}{\hfill\noindent{\lower.38cm\hbox{\rule{0.2pt}{0.4cm}}\rule{8.35cm}
              {0.2pt}}\vspace*{-0.5cm}}

\let\asas=\cite
\renewcommand\cite[1]{\raisebox{0.3mm}{\textsuperscript{\asas{#1}}}}

\makeatletter

\renewcommand{\thefootnote}{\fnsymbol{footnote}}

\renewcommand{\thanks}[1]{\thanksmark
    \protected@xdef\@thanks{\@thanks
        \protect\footnotetext[0]{\hspace*{-6pt}$*$\;#1}}}

\newcounter{email}
\newcommand{\email}[1]{%
    \protected@xdef\@thanks{\@thanks%
        \protect\footnotetext[0]{\hspace*{-8pt}\arabic{email})\,{E-mail:\,}#1}}%
        \stepcounter{email}}%

\renewcommand\footnoterule{
  \kern 1\p@
  \hrule \@width37mm
  \kern 8\p@}

\renewcommand\@makefntext[1]{%
    \parindent 1em%
    \noindent
    \hb@xt@2em{\hss\@makefnmark}#1}

\renewcommand\maketitle{\par
  \begingroup
    \renewcommand\thefootnote{\@fnsymbol\c@footnote}%
    \def\@makefnmark{\rlap{\@textsuperscript{\normalfont\@thefnmark}}}%
    \long\def\@makefntext##1{\parindent 1em\noindent
            \hb@xt@2em{%
                \hss\@textsuperscript{\normalfont\@thefnmark}}##1}%
    \if@twocolumn
      \ifnum \col@number=\@ne
        \@maketitle
      \else
        \twocolumn[\@maketitle]%
      \fi
    \else
      \newpage
      \global\@topnum\z@   
      \@maketitle
    \fi
  \@thanks
  \endgroup
  \setcounter{footnote}{0}%
  \global\let\thanks\relax
  \global\let\maketitle\relax
  \global\let\@maketitle\relax
  \global\let\@thanks\@empty
  \global\let\@author\@empty
  \global\let\@date\@empty
  \global\let\@title\@empty
  \global\let\title\relax
  \global\let\author\relax
  \global\let\date\relax
  \global\let\and\relax}

\renewcommand\@maketitle{%
  \begin{center}%
  \let \footnote \thanks
   \vspace*{0.5em}
    {\LARGE\bf \@title \par}%
    {\normalsize%
      \lineskip .5em%
      \vskip 2em%
      \begin{tabular}[t]{c}%
        \@author%
      \end{tabular}}%
  \end{center}}%
\makeatother

\newcommand{\danwei}[1]{%
  \begin{center}%
    \vskip -1em%
    \begin{center}%
      {\footnotesize #1}%
    \end{center}%
  \end{center}%
}%

\renewenvironment{abstract}%
  {\small\vspace{0.5mm}%
   \list{}{\rightmargin 2em%
           \leftmargin 2em}%
    \item{}{\bf Abstract}\hspace*{0.5em}\relax}%
   {\endlist}

\newenvironment{keyword}%
  {\small\vspace{1mm}%
    \list{}{\rightmargin 2em%
           \leftmargin 2em}%
                \item{}{\bf Key~words}\hspace*{0.5em}\relax }%
       {\endlist%
        }%

  {\small%
    \list{}{\rightmargin 2em%
           \leftmargin 2em}%
                \item{}{\bf PACS}\hspace*{0.5em}\relax }%
       {\endlist%
        \vskip 6mm}%

\makeatletter

\renewcommand \thesection {\bf\@arabic\c@section}

\renewcommand\section{\@startsection {section}{1}{\z@}%
                                    {5mm \@plus.2ex \@minus .2ex}%
                                   {5mm \@plus.2ex \@minus .2ex}%
                                   {\normalfont\large\bfseries}}
\renewcommand\subsection{\@startsection{subsection}{2}{\z@}%
                                     {1.5ex \@plus .2ex}%
                                     {1.5ex \@plus .2ex}%
                                     {\normalfont\bfseries}}

\renewcommand\subsubsection{\renewcommand \thesection {\@arabic\c@section}
                            \@startsection{subsubsection}{3}{\z@}%
                                     {0.5ex}%
                                     {0.5ex}%
                                     {\normalfont}}

\renewcommand{\@biblabel}[1]{#1}
\renewcommand\refname{{\normalsize\bf References}}
\renewenvironment{thebibliography}[1]
     {\noindent\refname%
      \@mkboth{\MakeUppercase\refname}{\MakeUppercase\refname}%
      \footnotesize
      \list{\@biblabel{\@arabic\c@enumiv}}%
           {\settowidth\labelwidth{\@biblabel{#1}}%
            \leftmargin\labelwidth
            \advance\leftmargin\labelsep
            \@openbib@code
            \usecounter{enumiv}%
            \let\p@enumiv\@empty
            \renewcommand\theenumiv{\@arabic\c@enumiv}}%
      \setlength{\itemsep}{0mm}
      \setlength{\labelsep}{0.8em}
      \setlength{\parsep}{0mm}
      \setlength{\parskip}{0mm}
      \setlength{\topsep}{0mm}
      \setlength{\partopsep}{0mm}
      \clubpenalty4000
      \@clubpenalty \clubpenalty
      \widowpenalty4000%
      \sfcode`\.\@m}
     {\def\@noitemerr
       {\@latex@warning{Empty `thebibliography' environment}}%
      \endlist}

\newenvironment{mylabc}
                {%
                 \newsmall
                 \let\\\@centercr
                 \list{}{\itemsep      \z@
                         \itemindent   -1em%
                         \listparindent0em
                         \leftmargin   3em
                         \rightmargin  2em}
                         \item\relax}
                {\endlist}

                {\footnotesize
                 \vspace{-1mm}
                 \let\\\@centercr
                 \list{}{\itemsep      \z@
                         \itemindent   0em%
                         \listparindent0em
                         \leftmargin   2mm
                         \rightmargin  0em}
                         \item\relax}

                {\endlist}
\makeatother

\makeatletter

\renewcommand\caption[1]{%
\sbox\@tempboxa{\newsmall #1}%
\ifdim \wd\@tempboxa >\hsize
\begin{mylabc}
\vspace{-2mm}
{\small #1}%
\vskip 1mm%
\end{mylabc}
\else
\global \@minipagefalse
\vspace*{-2mm}
\hb@xt@\hsize{\hfil\box\@tempboxa\hfil}%
\vskip 1mm%
\fi}
\makeatother

\begin{document}

\footnotetext[0]{Received 8 September 2005,\quad Revised 25 October 2005}

\title{On Statistical Significance of Signal\thanks{Supported by National Natural Science Foundation
of China(19991483)}}

\author{%
ZHU Yong-Sheng\emark{1)}\email{zhuys@ihep.ac.cn}%
}

\maketitle

\danwei{%
(Institute of High Energy Physics, CAS, Beijing 100049, China)\\
}

\begin{abstract}
A definition for the statistical significance of a signal
in an experiment is proposed by establishing a correlation between the
observed $p$-value and the normal distribution integral probability, which is
suitable for both counting experiment and continuous test statistics. The
explicit expressions to calculate the statistical significance for both
cases are given.
\end{abstract}


\begin{multicols}{2}

\section{Introduction}

The statistical significance of a signal in an experiment of particle
physics is to quantify the degree of confidence that the observation in the
experiment either confirm or disprove a null hypothesis $H_0$, in favor of
an alternative hypothesis $H_1$. Usually $H_0$ stands for the known or
background processes, while the alternative hypothesis $H_1 $ stands for a
new or a signal process plus background processes with respective production
cross section. This concept is very useful for usual measurements that one
can have an intuitive estimation, to what extent one can believe the
observed phenomena are due to backgrounds or a signal. It becomes crucial
for the measurements which claim a new discovery or a new signal. As a
convention in particle physics experiment, the ``5$\sigma $'' standard,
namely the statistical significance $S\geqslant 5$ is required to define the
sensitivity for discovery; while in the cases $S\geqslant 3$ ($S\geqslant 2)$, one may
claim that the observed signal has strong (weak) evidence.

However, as pointed out in Ref. [1], the concept of the statistical
significance has not been employed consistently in the most important
discoveries made over the last quarter century. Also, the definitions of the
statistical significance in different measurements differ from each other.
Listed below are various definitions for the statistical significance in
counting experiment (see, for example, Refs. [2---4]):
\begin{equation}
\label{eq1}
S_1 =(n-b)/\sqrt b ,
\end{equation}
\begin{equation}
\label{eq2}
S_2 =(n-b)/\sqrt n ,
\end{equation}
\begin{equation}
\label{eq3}
S_{12} =\sqrt n -\sqrt b ,
\end{equation}
\begin{equation}
\label{eq4}
S_{B1} = S_1 -k(\alpha )\sqrt {n / b} ,
\end{equation}
\begin{equation}
\label{eq5}
S_{B12} =2S_{12} - k(\alpha ),
\end{equation}
\begin{equation}
\label{eq6}
\vint\nolimits_{-\infty }^{S_N } {N(0,1)\dao x} =\sum\limits_{i=0}^{n-1}
{{\rm e}^{-b}\frac{b^i}{i!}} ,
\end{equation}
where $n$ is the total number of the observed events, which is the Poisson
variable with the expectation $s+b$, $s$ is the expected number of signal
events to be searched, while $b$ is the known expected number of Poisson
distributed background events. All numbers are counted in the ``signal
region'' where the searched signal events are supposed to appear. In
Eqs. (\ref{eq4}) and (\ref{eq5}), $k(\alpha )$ is a factor related to $\alpha $ that the
corresponding statistical significance assumes $1- \alpha $ acceptance for
positive decision about signal observation, and
$k(0.5)=0$, $k(0.25)=0.66$, $k(0.1)=1.28$, $k(0.05)=1.64$ etc.\cite{3}. In Eq. (\ref{eq6}),
$N(0,1)$ is a notation for the normal function with the expectation and
variance equal to 0 and 1, respectively. On the other hand, the measurements
in particle physics often examine statistical variables that are continuous
in nature. Actually, to identify a sample of events enriched in the signal
process, it is often important to take into account the entire distribution
of a given variable for a set of events, rather than just to count the
events within a given signal region of values. In this situation, I. Nasky\cite{4}
gives a definition of the statistical significance via likelihood function
\begin{equation}
\label{eq7}
S_L =\sqrt {-2\ln L(b)/L(s+b)}
\end{equation}
under the assumption that $-2\ln L(b)/L(s+b)$ distributes as $\chi ^2$
function with degree of freedom of 1.

Upon the above situation, it is clear that we desire to have a
self-consistent definition for statistical significance, which can avoid the
danger that the same $S$ value in different measurements may imply virtually
different statistical significance, and can be suitable for both counting
experiment and continuous test statistics. In this letter we propose a
definition of the statistical significance, which could be more close to the
desired property stated above.

\section{Definition of the statistical significance}

The $p$-value is defined to quantify the level of agreement between the
experimental data and a hypothesis\cite{1,5}. Assume an experiment
makes a measurement for test statistic $t$ being equal to $t_{\rm obs}$, and
$t$ has a probability density function $g(t\vert H_0 )$ if a null hypothesis
$H_0$ is true. We futher assume that large $t$ values correspond to poor
agreement between the data and the null hypothesis $H_0 $, then the
$p$-value of an experiment would be
\begin{equation}
\label{eq8}
p(t_{\rm obs} )=P(t>t_{\rm obs} \vert H_0 )= \vint\nolimits_{t_{\rm obs}}^\infty {g(t\vert H_0)\dao t} .
\end{equation}
A very small $p$-value tends to reject the null hypothesis $H_0$.

Since the $p$-value of an experiment provides a measure of the consistency
between the $H_0 $ hypothesis and the measurement, our definition for
statistical significance $S$ relates with the $p$-value in the form of
\begin{equation}
\label{eq9}
\vint\nolimits_{-S}^S {N(0,1)\dao x = 1-p(t_{\rm obs})},
\end{equation}
under the assumption that the null hypothesis $H_0 $ represents that the
observed events can be described merely by background processes. Because a
small $p$-value means a small probability of $H_0 $ being true, corresponds to
a large probability of $H_1 $ being true, one would get a large signal
significance $S$ for a small $p$-value, and vice versa. The left side of
Eq. (\ref{eq9}) represents the probability of the normal distribution in the
region within $\pm S$ standard deviation ($\pm S\sigma )$, therefore,
this definition conforms itself to the meaning of that the statistical
significance should have. In such a definition, some correlated $S$ and
$p$-values are listed in Table 1.

\begin{center}
\caption{Table~1.\quad The statistical significance $S$ and
correlated $ p$-value.}
\footnotesize
\begin{tabular*}{80mm}{c@{\extracolsep{\fill}}ccc}
\toprule
&$S$  & $p$-value &\\
\hline
&1 & 0.3173 &\\
&2 & 0.0455 &\\
&3 & 0.0027 &\\
&4 & 6.3$\times$10$^{-5}$ &\\
&5 & 5.7$\times$10$^{-7}$ &\\
&6 & 2.0$\times$10$^{-9}$ &\\
\bottomrule
\end{tabular*}\label{tab1}
\vspace{2mm}
\end{center}

\section{Statistical significance in counting experiment}

\vspace{-2mm}
A group of particle physics experiment involves the search for new phenomena
or signal by observing a unique class of events that can-not be described by
background processes. One can address this problem to that of a ``counting
experiment'', where one identifies a class of events using well-defined
criteria, counts up the number of observed events, and estimates the average
rate of events contributed by various backgrounds in the signal region,
where the signal events (if exist) will be clustered. Assume in an
experiment, the number of signal events in the signal region is a Poisson
variable with the expectation $s$, while the number of events from
backgrounds is a Poisson variable with a known expectation $b$ without
error, then the observed number of events distributes as the Poisson
variable with the expectation $s+b$. If the experiment observed $n_{\rm obs}$
events in the signal region, then the $p$-value is
\begin{eqnarray}
\label{eq10}
p(n_{\rm obs} ) &=& P(n>n_{\rm obs} \vert H_0 ) = \nonumber\\
&& \sum\limits_{n=n_{\rm obs} }^\infty
{\frac{b^n}{n!}{\rm e}^{-b} = 1 - \sum\limits_{n=0}^{n_{\rm obs} -1}
{\frac{b^n}{n!}{\rm e}^{-b}} }~.
\end{eqnarray}
Substituting this relation to Eq. (\ref{eq9}), one immediately has
\begin{equation}
\label{eq11}
\vint\nolimits_{-S}^S {N(0,1)\dao x} =\sum\limits_{n=0}^{n_{\rm obs} -1} {\frac{b^n}{n!}}
{\rm e}^{-b}~.
\end{equation}
Then, the signal statistical significance $S$ can be easily determined.
Comparing this equation with Eq. (\ref{eq6}) given by Ref. [4], we notice the
lower limit of the integral is different.

\section{Statistical significance in continuous test statistics}

The general problem in this situation can be addressed as follows. Suppose
we identify a class of events using well-defined criteria, which are
characterized by a set of $N$ observations $X_1,X_2,\cdots, X_N$ for a random
variable $X$. In addition, one has a hypothesis to test that predicts the
probability density function of $X$, say $f(X\vert \bm {\theta }),$ where
$\bm {\theta }=(\theta _1 ,\theta _2 ,\cdots , \theta _k )$ is a set of
parameters which need to be estimated from the data. Then the problem is to
define a statistic that gives a measure of the consistency between the
distribution of data and the distribution given by the hypothesis.

To be concrete, we consider the random variable $X$ is, say, an invariant
mass, and the $N$ observations $X_1 ,X_2 ,\cdots, X_N $ give an experimental
distribution of $X$. Assuming parameters $\bm {\theta }=(\theta _1 ,\theta
_2 , \cdots, \theta _k )\equiv (\bm {\theta }_s;\bm{\theta}_b)$, where
$\bm {\theta }_s $ and $\bm {\theta }_b $ represent the parameters related
to signal (say, a resonance) and backgrounds contribution, respectively. We
assume the null hypothesis $H_0 $ stands for that the experimental
distribution of $X$ can be described merely by the background
processes, while the alternative hypothesis $H_1 $ stands for that
the experimental distribution of $X$ should be described by the backgrounds
plus signal; namely, the null hypothesis $H_0 $ specifies the fixed value(s)
for a subset of parameters $\bm {\theta }_s $ (the number of fixed
parameter(s) is denoted as $r$), while the alternative hypothesis $H_1 $
leaves the $r$ parameter(s) free to take any value(s) other than those
specified in $H_0 $. Therefore, the parameters $\bm {\theta }$ are
restricted to lie in a subspace $\omega $ of its total space $\varOmega$. On
the basis of a data sample of size $N$ from $f(X\vert \bm {\theta }),$ we
want to test the hypothesis $H_0 : \bm {\theta}$ belongs to $\omega $.
Given the observations $X_1 ,X_2 ,\cdots, X_N $, the likelihood function is
$L=\prod\limits_{i=1}^N {f(X_i \vert \bm {\theta })}$. The maximum of this
function over the total space $\varOmega $ is denoted by $L(\hat{\varOmega })$;
while within the subspace $\omega $ the maximum of the likelihood function
is denoted by $L(\hat{\omega })$, then we define the likelihood-ratio
$\lambda \equiv L(\hat{\omega})/ L(\hat{\varOmega})$. It can be shown that for $H_0 $ true, the statistic
\begin{equation}
\label{eq12}
t\equiv -2\ln \lambda \equiv 2(\ln L_{\max } (s+b)-\ln L_{\max } (b))
\end{equation}
is distributed as $\chi ^2(r)$ when the sample size $N$ is large\cite{6}. In
Eq. (\ref{eq12}) we use $L_{\max }(s+b)$ and $L_{\max }(b)$ denoting $L(\hat{\varOmega })$
and $L(\hat {\omega })$, respectively. If $\lambda $ turns out
to be in the neighborhood of 1, the null hypothesis $H_0 $ is such that it
renders $L(\hat {\omega })$ close to the maximum $L(\hat {\varOmega })$, and
hence $H_0 $ will have a large probability of being true. On the other hand,
a small value of $\lambda $ will indicates that $H_0 $ is unlikely.
Therefore, the critical region of $\lambda $ is in the neighborhood of 0,
corresponding to a large value of statistic $t$. If the measured value of
$t$ in an experiment is $t_{\rm obs} $, from Eq. (\ref{eq8}) we have $p$-value
\begin{equation}
\label{eq13}
p(t_{\rm obs} )=\vint\nolimits_{t_{\rm obs} }^\infty {\chi ^2(t;r)\dao t} .
\end{equation}
Therefore, in terms of Eq. (\ref{eq9}), we can calculate the signal
significance according to the following expression:
\begin{equation}
\label{eq14}
\vint\nolimits_{-S}^S {N(0,1)\dao x = 1-p(t_{\rm obs} )} =\vint\nolimits_0^{\,t_{\rm obs} } {\chi ^2(t;r)\dao t} .
\end{equation}
For the case of $r=1$, we have
\[
\vint\nolimits_{-S}^{\,S} {N(0,1)} \dao x = \vint\nolimits_0^{\,t_{\rm obs} } {\chi ^2(t;1)\dao t =
2\vint\nolimits_0^{\,\sqrt{t_{\rm obs} } } {N(0,1)\dao x} } ,
\]
and immediately obtain
\begin{equation}
\label{eq15}
S = \sqrt {t_{\rm obs} } =[2(\ln L_{\max } (s+b)-\ln L_{\max }
(b))]^{1 /2},
\end{equation}
which is identical to Eq. (\ref{eq7}) given by Ref. [4].

\section{Discussion and summary}

In Section 2, the $p$-value defined by Eq. (\ref{eq8}) is based on the assumption
that large $t$ values correspond to poor agreement between the null
hypothesis $H_0 $ and the observed data, namely, the critical region of
statistic $t$ for $H_0 $ lies on the upper side of its distribution. If the
situation is such that the critical region of statistic $t$ lies on the
lower side of its distribution, then Eq. (\ref{eq8}) should be replaced by
\begin{equation}
\label{eq16}
p(t_{\rm obs} )=P(t<t_{\rm obs} \vert H_0 )=\vint\nolimits_{-\infty }^{t_{\rm obs} }
{g(t\vert H_0 )\dao t} ,
\end{equation}
and the definition of statistical significance $S$ expressed by Eq. (\ref{eq9})
is still applicable. For the case that the critical region of statistic $t$
for $H_0 $ lies on both lower and upper tails of its distribution, and one
determined from an experiment the observed $t$ values in both sides:
$t_{\rm obs}^U $ and $t_{\rm obs}^L $, then Eq. (\ref{eq8}) should be replaced by
\begin{eqnarray}
\label{eq17}
p(t_{\rm obs}) &=& P(t<t_{\rm obs}^L \vert H_0 ) + P(t>t_{\rm obs}^U \vert H_0)= \nonumber\\
&& \vint\nolimits_{-\infty }^{t_{\rm obs}^L } {g(t\vert H_0 )\dao t + \vint\nolimits_{t_{\rm obs}^U }^\infty
{g(t\vert H_0 )\dao t} } .
\end{eqnarray}

In summary, we proposed a definition for the statistical significance by
establishing a correlation between the normal distribution integral
probability and the $p$-value observed in an experiment, which is suitable for
both counting experiment and continuous test statistics. The explicit
expressions to calculate the statistical significance for counting
experiment and continuous test statistics in terms of the Poisson
probability and likelihood-ratio are given.
\end{multicols}

\vspace{-2mm}
\centerline{\rule{80mm}{0.1pt}}
\vspace{2mm}

\begin{multicols}{2}


\end{multicols}

\clearpage

\end{document}